Journal of
**Risk and Financial Management**

MDPI

*Article*

# Exploring Dynamic Asset Pricing within Bachelier's Market Model

**Nancy Asare Nyarko** [1,*], **Bhathiya Divelgama** [1], **Jagdish Gnawali** [1], **Blessing Omotade** [1], **Svetlozar T. Rachev** [1] and **Peter Yegon** [1]

1   Department of Mathematics and Statistics, Texas Tech University, Lubbock, TX 79409-1042, USA; bdivelga@ttu.edu (B.D.); jgnawali@ttu.edu (G.J.); bomotade@ttu.edu (B.O.); zari.rachev@ttu.edu (S.T.R.); pyegon@ttu.edu (P.Y.)
*   Correspondence: nasareny@ttu.edu

**Abstract:** This paper delves into the dynamics of asset pricing within Bachelier's market model (BMM), elucidating the representation of risky asset price dynamics and the definition of riskless assets. It highlights the fundamental differences between BMM and the Black-Scholes-Merton market model (BSMMM), including the extension of BMM to handle assets yielding a simple dividend. Our investigation further explores Bachelier's term structure of interest rates (BTSIR), introducing a novel version of Bachelier's Heath-Jarrow-Morton model and adapting the Hull-White interest rate model to fit BMM. The study concludes by examining the applicability of BMM in real-world scenarios, such as those involving environmental, social, and governance (ESG)-adjusted stock prices and commodity spreads.

## 1. Introduction

Louis Bachelier's groundbreaking option pricing model, presented in his Ph.D. thesis (Bachelier 1900), marked a significant milestone in the emergence of mathematical finance (Courtault et al. 2000; Davis and Etheridge 2006; Sullivan and Weithers 1991). Bachelier's model employed arithmetic Brownian motion to describe the price dynamics, and although it faced initial skepticism as a suitable asset price model, there were a few exceptional cases in which it proved to be applicable. Bachelier's market model necessitated the inclusion of a riskless bank account with a simple riskless interest rate, expanding upon Shiryaev's prior work on Bachelier's model with zero interest rates (Shiryaev 2003). However, we assert that combining Bachelier's asset dynamics, governed by arithmetic Brownian motion, with a riskless bank account employing compound interest rates, as explored by Terakado (2019) and Brooks and Brooks (2017), is not appropriate. The criticism regarding the fact that Bachelier's market model allows negative riskless rates has become irrelevant, given the existence of such instances in real financial markets (Neufeld 2022).

Presently, the primary challenge in utilizing Bachelier's market model arises from the requirement of non-negative asset prices. While stock prices inherently meet this criterion, the environmental, social, and governance (ESG)-adjusted price may exhibit negative values. In Rachev et al. (2023), we define the ESG-adjusted stock price $S_t^{ESG}$, $t \geq 0$, as follows:

$$S_t^{ESG} = S_t^{(X)}(1 + \gamma^{ESG} Z_t^{(X;I)}) \in R,$$

where $S_t^{(X)} > 0$ represents the stock price of company $X$, $Z_t^{(X;I)} \in R$ represents the relative ESG score of $X$, and $\gamma^{ESG} \in R$ denotes the ESG affinity of the financial market. The ESG-adjusted stock price introduces a crucial third dimension (ESG) into the traditional two-dimensional (risk-return) investment framework, as discussed by Lauria et al. (2022) and Hu et al. (2022). Some experts in the financial industry and financial regulators currently consider the ESG performance as an independent factor alongside the risk and return, which are the other two factors (Hale 2023).

Further instances of price dynamics that have positive and negative values encompass various scenarios. These include the disparity between two stock prices, the distinction between a stock price and a tradable benchmark security (such as a stock index or US Treasury securities), the variances in currency pairs and futures and option contracts, and the variations in the yield differentials of stocks or bonds (Schaefer 2002; Vaidya 2023). Additionally, commodity spreads entail dissimilarities in prices between different commodities, disparities in the prices of the same commodity across various trading locations, and variations in a commodity's prices at different stages of the production process. Noteworthy examples include crack spreads in the oil market, crush spreads in the soybean market, spark spreads in the electricity market, calendar spreads that indicate the prices of one commodity at different points in time, and so on (Downey 2022).

In this paper, we present the fundamentals of dynamic asset pricing within Bachelier's market model (BMM). In BMM, the risky asset price follows an arithmetic Brownian motion with a time-dependent instantaneous mean and variance that thus represents a general diffusion process. The novelty lies in the definition of the riskless asset, which, diverging from the Black-Scholes-Merton riskless bank account, is defined as a simple interest bank account. The necessity of selecting a simple interest bank account as the riskless asset was elucidated in Rachev et al. (2023).

Our exploration of BMM commences with the derivation of Bachelier's partial differential equation (PDE) using standard no-arbitrage arguments in a complete market model. Although Bachelier's PDE substantially differs from the Black-Scholes-Merton PDE, it is a particular form of the Cauchy PDE. Therefore, its solution is provided by the Feynman-Kac probabilistic solution to the Cauchy problem, as outlined in Duffie (2001, Appendix E). The solution of Bachelier's PDE furnishes the formula for European contingent claims (ECCs) within BMM. The continuous diffusion propelling the Feynman-Kac solution represents Bachelier's risk-neutral process, which we further elucidate through the risk-neutral valuation of ECCs in BMM.

We posit that BMM should not be merely viewed as a variant of the Black-Scholes-Merton market model (BSMMM) that follows an exponential transformation of the price dynamics of both risky and riskless assets. This position underscores the fundamental differences between BMM and the BSMMM. Additionally, we extend BMM to situations in which trading the risky asset yields a simple dividend. This definition of the simple dividend yield mirrors that of the simple interest rate associated with Bachelier's riskless asset. Yet again, the corresponding Bachelier's PDE is a special form of the Cauchy equation, with its probabilistic solution provided by the Feynman-Kac formula. Notably, we derive the no-arbitrage price of the zero-coupon bond within BMM, and this serves as a springboard for studying Bachelier's term structure of interest rates (BTSIR).

In the second half of the paper, we delve into the study of BTSIR. We derive both the no-arbitrage Bachelier's T-forward (delivery) price and Bachelier's future price. Attempting to define Bachelier's forward rates in a way that mirrors the definition of forward rates in the BSMMM proves somewhat unsatisfactory. It leads to the Heath-Jarrow-Morton (HJM) interest rate model for BMM, which has the same deficiencies as the HJM model for the BSMMM, triggering the introduction of the London Inter-Bank Offered Rate (LIBOR) model. We propose a new version of Bachelier's HJM model in which the forward rates are replaced with zero-simple-interest loans, more succinctly referred to as no-interest loans. This modification results in a new form of BTSIR defined by the dynamics of the no-interest loans. Finally, we adapt the Hull-White interest rate model to fit BMM.

This paper contains three main sections (subdivided into numerous subsections) and a conclusion. The current section serves as an introduction to BMM and outlines the principal findings of this study. Section 2 is dedicated to option pricing in BMM for assets with time-dependent parameters. Subsection 2.1 involves the derivation of Bachelier's PDE and the provision of its Feynman-Kac probabilistic solution, which leads to the no-arbitrage option price formula. In subsection 2.2, the same result is achieved using risk-neutral valuation in BMM. Subsection 2.3 explains why BMM and the BSMMM are fundamentally different. Subsection 2.4 introduces and studies Bachelier's option pricing model for assets with a simple dividend yield. Subsection 2.5 extends BMM to a risky asset with a dividend yield. Section 3 focuses on BTSIR. In subsection 3.1, we provide the risk-neutral valuation of Bachelier's zero-coupon bond and thus introduce BTSIR. In subsection 3.2, we define Bachelier's forward contract, and in subsection 3.3, we present Bachelier's future price. In subsection 3.4, we introduce Bachelier's forward rates and no-interest loan rates, and the HJM model for BMM. Finally, subsection 3.5 is devoted to the Hull and White model in BMM. Section 4 contains the concluding remarks.

## 2. Approaches to Option Pricing in Bachelier's Model

### 2.1. Bachelier's Option Pricing: The Partial Differential Equation and its Feynman-Kac Solution

Bachelier's market model $(\mathcal{A}, \mathfrak{B}, \mathcal{C})$ is defined by a risky asset $\mathcal{A}$, riskless asset $\mathfrak{B}$, and EEC[1] $\mathcal{C}$. $\mathcal{A}$ has the **Bachelier's asset price dynamics** of a continuous diffusion process determined by the stochastic differential equation (SDE)

$$dA_t = \rho_t dt + v_t dB_t, t \geq 0, \ A_0 > 0, \tag{1}$$

where $B_t, t \in [0, \infty)$, is a standard Brownian motion on a stochastic basis (filtered probability space) $(\Omega, \mathcal{F}, \mathbb{F} = \{\mathcal{F}_t = \sigma(B_u, u \leq t) \subseteq \mathcal{F}, t \geq 0\}, \mathbb{P})$ on a complete probability space $(\Omega, \mathcal{F}, \mathbb{P})$, $\rho_t = \rho(A_t, t)$ and $v_t = v(A_t, t), t \geq 0$, are $\mathbb{F}$-adapted processes, and $\rho \colon \mathbb{R} \times [0, \infty) \to \mathbb{R}$ and $v \colon \mathbb{R} \times [0, \infty) \to (0, \infty)$ satisfy the usual regularity conditions[2]. From (1), we have

$$A_t = A_t + \int_0^t \rho_s ds + \int_0^t v_s dB_s, t \geq 0. \tag{2}$$

$\mathfrak{B}$ is the **simple interest bank account**[3] with price dynamics determined by the SDE

$$d\beta_t = r_t dt, \beta_0 > 0, \tag{3}$$

where **the simple interest rate** $r_t = r(A_t, t), t \geq 0$, is an $\mathbb{F}$-adapted process, $r \colon \mathbb{R} \times [0, \infty) \to \mathbb{R}$ satisfies the usual regularity conditions[4], and $r_t \leq \rho_t, t \geq 0$ ($\mathbb{P}$-almost surely ($\mathbb{P}$-a.s.)).
From (3), we have

---

[1] An EEC is also called a European derivative security (shortly, derivative) or European option (shortly, option).

[2] See Duffie (2001, Section 5G and Appendix E). The regularity conditions are satisfied if $\rho(x, t), x \in \mathbb{R}, t \geq 0$, and $v(x, t), x \in \mathbb{R}, t \geq 0$, are measurable and satisfy the Lipschitz and growth conditions in $x \in \mathbb{R}$ and $\int_0^t |\rho_s| ds < \infty, \int_0^t v_s^2 ds < \infty$ for all $t \geq 0$. To simplify the exposition, we will assume that $\rho_t, t \geq 0$, and $v_t, t \geq 0$, have trajectories that are continuous and uniformly bounded on $[0, \infty)$.

[3] See Rachev et al. (2023). One can view $\mathfrak{B}$ as a **piggy bank account**; see Sallie Mae (2023). If $\beta_0 = 0, r_t = prt$, then $\mathfrak{B}$ is a simple interest savings bank account; see Burnette (2022). We prefer to call $\mathfrak{B}$ a simple interest bank account rather than a piggy bank account, following the definition of $\beta_t, t \geq 0$, given in (3) and (4).

[4] The regularity conditions are satisfied if $r(x, t), x \in \mathbb{R}, t \geq 0$, is measurable and $\int_0^t |r_s| ds < \infty$ for all $t \geq 0$. To simplify the exposition, we will assume that $r_t, t \geq 0$, has trajectories that are continuous and uniformly bounded on $[0, \infty)$.

$$\beta_t = \beta_0 + \int_0^t r_s \, ds. \tag{4}$$

The EEC $\mathcal{C}$ has the price dynamics

$$C_t = f(A_t, t), t \in [0, T], \tag{5}$$

where $f(x, t), x \in R, t \in [0, T]$, has continuous partial derivatives $\frac{\partial^2 f(x,t)}{\partial x^2}, \frac{\partial f(x,t)}{\partial t}$ on $t \in [0, T)$, and $T$ is the terminal (expiration) time of $\mathcal{C}$, with the EEC's terminal payoff $C_T = g(A_T)$ for some continuous function $g: \mathbb{R} \to \mathbb{R}$.

Next, we will derive Bachelier's PDE for $f(x, t), x \in R, t \in [0, T)$. We start with the Itô formula[5] for $C_t = f(A_t, t)$:

$$dC_t = \left[ \frac{\partial f(A_t, t)}{\partial t} + \frac{\partial f(A_t, t)}{\partial x} \rho_t + \frac{1}{2} \frac{\partial^2 f(A_t, t)}{\partial x^2} v_t^2 \right] dt + \frac{\partial f(A_t, t)}{\partial x} v_t \, dB_t. \tag{6}$$

Consider a self-financing riskless portfolio of $\mathcal{A}$ and $\mathcal{C}$, which can be written as $P_t = a_t A_t + C_t$. Thus, $dP_t = a_t dA_t + dC_t = d\beta_t$, which leads to

$$a_t(\rho_t dt + v_t dB_t)$$

$$+ \left[ \frac{\partial f(A_t, t)}{\partial t} + \frac{\partial f(A_t, t)}{\partial x} \rho_t + \frac{1}{2} \frac{\partial^2 f(A_t, t)}{\partial x^2} v_t^2 \right] dt + \frac{\partial f(A_t, t)}{\partial x} v_t \, dB_t = r_t dt.$$

Thus, $a_t = -\frac{\partial f(A_t, t)}{\partial x}$, and

$$a_t \rho_t + \left[ \frac{\partial f(A_t, t)}{\partial t} + \frac{\partial f(A_t, t)}{\partial x} \rho_t + \frac{1}{2} \frac{\partial^2 f(A_t, t)}{\partial x^2} v_t^2 \right] = r_t.$$

This leads to $\frac{\partial f(A_t, t)}{\partial t} + \frac{1}{2} \frac{\partial^2 f(A_t, t)}{\partial x^2} v_t^2 - r_t = 0$.

**Bachelier's PDE** for $f(x, t), x \in R, t \in [0, T)$, is

$$\frac{\partial f(x, t)}{\partial t} + \frac{1}{2} \frac{\partial^2 f(x, t)}{\partial x^2} v(x, t)^2 - r(x, t) = 0, \tag{7}$$

with the boundary condition $f(x, T) = g(x), x \in R$.

Next, we provide the Feynman-Kac solution of (7)[6]. First, let us recall the **Cauchy problem**: *Find $f \in \mathbb{C}^{2,1}(\mathbb{R} \times [0, T))$ by solving*

$$\mathfrak{D}f(x, t) - a(x, t)f(x, t) + h(x, t) = 0, x \in R, t \in [0, T), \tag{8}$$

*with the boundary condition $f(x, T) = g(x), x \in R$, where*

$$\mathfrak{D}f(x, t) = \frac{\partial f(x, t)}{\partial t} + \frac{\partial f(A(t), t)}{\partial x} \mu(x, t) + \frac{1}{2} \frac{\partial^2 f(x, t)}{\partial x^2} \sigma(x, t)^2, \tag{9}$$

*where $a: \mathbb{R} \times [0, T] \to \mathbb{R}, h: \mathbb{R} \times [0, T] \to \mathbb{R}, g: \mathbb{R} \to \mathbb{R}, \mu: \mathbb{R} \times [0, T] \to \mathbb{R}, and \ \sigma: \mathbb{R} \times [0, T] \to \mathbb{R}.$*

Let $\tilde{B}(t), t \geq 0$, be a Brownian motion, and consider the Itô process $X_t, t \geq 0$, with the SDE

$$dX_t = \mu(X_t, t)dt + \sigma(X_t, t)d\tilde{B}(t), t \geq 0, \tag{10}$$

---

[5] See Duffie (2001, Section 5D).

[6] See the Feynman-Kas solution of the Cauchy problem in Duffie (2001, Appendix E).

assuming that $\mu:\mathbb{R}\times[0,T]\to\mathbb{R}$ and $\sigma:\mathbb{R}\times[0,T]\to\mathbb{R}$ satisfy the regularity conditions, guaranteeing that (10) has a unique strong solution. The Feynman-Kac (probabilistic) solution of (7)–(9) is given by

$$f(x,t) = \mathbb{E}^{X_t=x}\left[\int_t^T \varphi_{t,s}h(X_s,s)ds + \varphi_{t,T}g(X_T)\right], \tag{11}$$

where $\varphi_{t,s} = \exp\left[-\int_t^s a(X_u,u)du\right]$.

The solution (11) solves (7), with $a(x,t) = 0, h(x,t) = -r(x,t), \ \mu(x,t) = 0$, and $\sigma(x,t) = v(x,t)$. Thus, the Feynman-Kac solution of Bachelier's PDE (7) is

$$f(x,t) = \mathbb{E}^{Z_t=x}\left[g(Z_T) - \int_t^T r(Z_s,s)ds\right], \tag{12}$$

where $Z_t, t \geq 0$, is the Itô process satisfying the SDE

$$dZ_t = v(Z_t,t)d\tilde{B}(t), t \geq 0. \tag{13}$$

The value $C_t, t \in [0,T]$, of the option contract $\mathcal{C}$ is given by $C_t = f(A_t,t)$, where $f(x,t)$, $x = A_t$, is determined by (12) and (13).

Bachelier's PDE for $f(x,t), x \in R, t \in [0,T]$, with a constant volatility $v(x,t) = v$ and a constant simple riskless rate $r(x,t) = r$, is

$$\frac{\partial f(x,t)}{\partial t} + \frac{1}{2}\frac{\partial^2 f(x,t)}{\partial x^2}v^2 - r = 0, t \in [0,T), \tag{14}$$

with a boundary condition $f(x,T) = g(x)$, which is a non-homogeneous heat equation in which the time $t$ has been replaced with the time to maturity $\tau = T - t$[7].

As in Shirvani et al. (2020), it will be of interest to find a perpetual derivative $\mathcal{C}^{(\gamma)}$ with a price process $C_t^{(\gamma)} = f(A_t,t) = A_t^2 + \gamma\beta_t$ for some $\gamma \in R \setminus \{1,2\}$. The parameter value $\gamma$ is the solution of (7), where $f(x,t) = x^2 + \gamma(\beta_0 + rt)$. From (7), we have

$$\frac{\partial f(x,t)}{\partial t} + \frac{1}{2}\frac{\partial^2 f(x,t)}{\partial x^2}v^2 - r = \gamma r + v^2 - r = 0,$$

and thus, $\gamma = 1 - \frac{v^2}{r}$. The market $\left(\mathcal{A},\mathfrak{B},\mathcal{C}^{(\gamma)}\right)$ is arbitrage-free and complete, and thus, any security can be replicated using the other two. For example, if a simple risk account $\mathfrak{B}$ is not available for trading (Rachev et al. 2017), the trader can use a self-financing portfolio of the asset $\mathcal{A}$ and the perpetual derivative $\mathcal{C}^{(\gamma)}$ to replicate $\mathfrak{B}$. $\mathcal{C}^{(\gamma)}$ will also be applied in the construction of Bachelier's trinomial tree, which is needed in BTSIR models.

*2.2. Risk-Neutral Valuation in Bachelier's Option Pricing*

Consider again BMM $(\mathcal{A},\mathfrak{B},\mathcal{C})$, where the asset price $A_t, t \geq 0$, is given by (1), the simple interest account is given by (3), and the option price $C_t, t \in [0,T]$, defined by (5) has a terminal payoff $C_T = g(A_T)$. We search for the unique equivalent martingale measure $\mathbb{Q} \sim \mathbb{P}$ such that on $\mathbb{Q}$,

$$dA_t = r_t dt + v_t dB_t^{(\mathbb{Q})}, \ \ t \in [0,T], \tag{15}$$

where $B_t^{(\mathbb{Q})}$ is a standard Brownian motion on $\mathbb{Q}$. On $\mathbb{P}$, $B_t^{(\mathbb{Q})}$ is an arithmetic Brownian motion $dB_t^{(\mathbb{Q})} = dB_t + \theta_t dt$. Then, by choosing $\theta_t = \frac{\rho_t - r_t}{v_t} > 0$ to be the market price of risk, we obtain the dynamics of $A_t, t \geq 0$, on $\mathbb{Q}$, as defined in (15). Thus, the excess asset price over the riskless simple interest rate,

---

[7] For the solution of the nonhomogeneous heat equation, see Herman (2023).

$$A_t - \beta_t = A_0 - \beta_0 + \int_0^t v_s dB_s^{(\mathbb{Q})}, \tag{16}$$

is a martingale on $(\Omega, \mathcal{F}, \mathbb{F}, \mathbb{Q})$. This is the main difference between BMM and the BSMMM, where the discounted price process under the risk-neutral measure is a martingale.

Using the same arguments used in Duffie (2001, Section 6H), we have that the option price in excess of the riskless simple interest rate is a martingale under $\mathbb{Q}$, that is,

$$C_t - \beta_t = E_t^{\mathbb{Q}}(C_T - \beta_T) = E_t^{\mathbb{Q}}(g(A_T) - \beta_T). \tag{17}$$

Thus,

$$C_t = E_t^{\mathbb{Q}}\left(g(A_T) - \int_t^T r_s ds\right) = E_t^{\mathbb{Q}}\left(g\left(A_t + \int_t^T r_s ds + \int_t^T v_s B_s^{(\mathbb{Q})}\right) - \int_t^T r_s ds\right). \tag{18}$$

Indeed, the option value $C_t$ in (18) and the value $C_t = f(A_t, t)$, where $f(x, t)$ is given in (12), coincide.

In the case when, in (18), $r_t = r$ and $v_t = v$, the price dynamics of a call option with a maturity $T$ and strike price $K$ are given by[8]

$$C_t = \lambda_{0,t} \Phi\left(\frac{\lambda_{0,t}}{\lambda_{1,t}}\right) + \lambda_{1,t} \varphi\left(\frac{\lambda_{0,t}}{\lambda_{1,t}}\right) - r(T - t), t \in [0, T], \tag{19}$$

where $\lambda_{0,t} = A_t - K + r(T - t)$, $\lambda_{1,t} = v\sqrt{T - t}$, $\Phi$ is the cumulative probability distribution of a standard normal random variable, and $\varphi = \Phi'$ is its probability density.

### 2.3. Distinguishing Bachelier's PDE from the Black-Scholes-Merton PDE

One can perceive that Bachelier's PDE (7) can be derived as a special case of the Black-Scholes-Merton PDE[9]. Indeed, in BMM, the risky asset follows the dynamics $dA_t = \rho_t dt + v_t dB_t, t \geq 0$, $A_0 > 0$ (see (1)), and the simple interest bank account dynamics are given by $d\beta_t = r_t dt, \beta_0 > 0$ (see (3)). It is tempting to consider the processes $S_t = e^{A_t}, t \geq 0$, and $b_t = e^{\beta_t}, t \geq 0$. Then, by Itô's lemma, $dS_t = \mu(S_t, t)dt + \sigma(S_t, t)dB_t$, where $\mu(S_t, t) = \left(\rho_t + \frac{1}{2}v_t^2\right)S_t$ and $\sigma(S_t, t) = v_t S_t$. Additionally, $db_t = r_t b_t dt$. Consider a Bachelier's ECC with a maturity $T$ and price process $C_t = f(A_t, t) = f(lnS_t, t) = h(S_t, t)$, where $h(x, t) = f(lnx, t), x > 0, t \in [0, T)$. The Black-Scholes-Merton PDE for $h(S_t, t)$ is

$$\frac{\partial h(x,t)}{\partial t} + r_t x \frac{\partial h(x,t)}{\partial x} + \frac{1}{2}\sigma(x,t)^2 \frac{\partial^2 h(x,t)}{\partial x^2} - r_t h(x,t) = 0, x > 0, t \in [0, T). \tag{20}$$

As $h(x, t) = f(lnx, t)$, (20) is equivalent to

$$\frac{\partial f(y,t)}{\partial t} + \left(r_t - \frac{1}{2}v_t^2\right)\frac{\partial f(y,t)}{\partial x} + \frac{1}{2}v_t^2 \frac{\partial^2 f(y,t)}{\partial x^2} - r_t f(y, t) = 0, \tag{21}$$

with $y = lnx \in R, t \in [0, T)$.

The Black-Scholes-Merton equation for $f(y, t) = f(lnx, t) = h(x, t))$, where $C_t = f(A_t, t) = f(lnS_t, t) = h(S_t, t)$, is different from Bachelier's PDE for $f(y, t)$, which is defined as follows (see (7)):

$$\frac{\partial f(y,t)}{\partial t} + \frac{1}{2}\frac{\partial^2 f(y,t)}{\partial x^2} v_t^2 - r_t = 0, y \in R, t \in [0, T). \tag{22}$$

The discrepancy arises from the fact that in the BSMMM, necessary arbitrage considerations are applied to the price process $S_t = \exp(A_t), t \geq 0$, and the bank account $b_t = e^{\beta t}, t \geq 0$. In contrast, in BMM, no-arbitrage arguments are applied, assuming that the asset price processes are the tradable securities, with the asset price process $A_t$ and the simple interest bank account $\beta_t, t \geq 0$. A holder of a short position in an ECC in BMM hedges her position using self-financing portfolios of the assets $\mathcal{A}$ and $\mathfrak{B}$, with dynamics $A_t$ and $\beta_t, t \geq 0$. The processes $S_t = e^{A_t}$ and $b_t = e^{\beta t}$ are not the price processes of traded assets in BMM. The processes $S_t = e^{A_t}$ and $b_t = e^{\beta t}$ cannot serve as the price processes of a traded security (a perpetual derivative) in Bachelier's no-arbitrage complete market. For $S_t = e^{A_t}, t \geq 0$, to be a perpetual derivative on Bachelier's market $(\mathcal{A}, \mathfrak{B})$, the function $f(x,t) = e^x$ should satisfy (7), which it does not. Therefore, in Bachelier's market, the process $S_t = e^{A_t}, t \geq 0$, cannot be utilized for hedging with a self-financing option replicating portfolio.

---

[8] See Theorem 1 in Rachev et al. (2023).

[9] See Duffie (2001, Section 5.G, p. 92, formula (23)).

## 2.4. Bachelier's Option Pricing Model for Assets with Simple Dividend Yield

Suppose that the asset $\mathcal{A}^{(\mathfrak{D})}$ pays dividends at a continuous rate [10] with an **instantaneous simple dividend yield** $\mathfrak{D}_t \in \mathbb{R}$, that is, in $(t, t+dt]$, the asset-holder receives $\mathfrak{D}_t dt$. Consider an EEC $\mathcal{C}$ in the BMM $(\mathcal{A}^{(\mathfrak{D})}, \mathfrak{B})$, with a price process $C_t^{(\mathfrak{D})}, t \in [0,T]$, where $T$ is the terminal time and the terminal payoff is $C_T^{(\mathfrak{D})} = g(\mathcal{A}_T^{(\mathfrak{D})})$. Therefore, the price process of $\mathcal{A}^{(\mathfrak{D})}$ is given by

$$dA_t^{(\mathfrak{D})} = \rho_t dt + v_t dB_t - \mathfrak{D}_t dt = (\rho_t - \mathfrak{D}_t)dt + v_t dB_t. \tag{23}$$

Consider a self-financing riskless portfolio of $\mathcal{A}$ and $\mathcal{C}$, which can be written as $P_t^{(\mathfrak{D})} = a_t A^{(\mathfrak{D})}(t) + C_t^{(\mathfrak{D})}$. Thus, $dP_t^{(\mathfrak{D})} = a_t dA_t^{(\mathfrak{D})} + dC_t^{(\mathfrak{D})} + a_t \mathfrak{D}_t dt = d\beta_t$, which leads to $a_t((\rho_t - \mathfrak{D}_t)dt + v_t dB_t) + dC_t^{(\mathfrak{D})} + a_t \mathfrak{D}_t dt = r_t dt$. From (23) and since $C_t^{(\mathfrak{D})} = f^{(\mathfrak{D})}(A_t^{(\mathfrak{D})}, t)$,

$$dC_t^{(\mathfrak{D})} = \left[ \frac{\partial f^{(\mathfrak{D})}(A_t^{(\mathfrak{D})}, t)}{\partial t} + \frac{\partial f^{(\mathfrak{D})}(A_t^{(\mathfrak{D})}, t)}{\partial x}(\rho_t - \mathfrak{D}_t) + \frac{1}{2}\frac{\partial^2 f^{(\mathfrak{D})}(A_t^{(\mathfrak{D})}, t)}{\partial x^2}v_t^2 \right] dt$$

$$+ \frac{\partial f^{(\mathfrak{D})}(A_t^{(\mathfrak{D})}, t)}{\partial x} v_t dB_t.$$

Thus, $a_t = -\frac{\partial f^{(\mathfrak{D})}(A_t^{(\mathfrak{D})}, t)}{\partial x}$, which leads to

$$\frac{\partial f^{(\mathfrak{D})}(A_t^{(\mathfrak{D})}, t)}{\partial t} - \frac{\partial f^{(\mathfrak{D})}(A_t^{(\mathfrak{D})}, t)}{\partial x}\mathfrak{D}_t + \frac{1}{2}\frac{\partial^2 f^{(\mathfrak{D})}(A_t^{(\mathfrak{D})}, t)}{\partial x^2}v_t^2 = r_t.$$

Thus, **Bachelier's PDE for an EEC on a stock paying dividends** with a rate $\mathfrak{D}_t$ is

$$\frac{\partial f^{(\mathfrak{D})}(x,t)}{\partial t} - \frac{\partial f^{(\mathfrak{D})}(x,t)}{\partial x}\mathfrak{D}_t + \frac{1}{2}\frac{\partial^2 f^{(\mathfrak{D})}(x,t)}{\partial x^2}v_t^2 - r_t = 0, f^{(\mathfrak{D})} = g(x). \tag{24}$$

We solve (24) using the Feynman-Kac solution ((12) and (13)) of the Cauchy problem ((8) and (9)), with $\mathbf{a(x,t) = 0}, \mathbf{\mu(x,t) = -\mathfrak{D}_t}, \mathbf{\sigma(x,t) = v_t}$, and $\mathbf{h(x,t) = -r_t}$. Thus, the Feynman-Kac solution of Bachelier's PDE **for an EEC on a stock paying dividends** with a rate $\mathfrak{D}_t$ (24) is

$$f^{(\mathfrak{D})}(x,t)(x,t) = \mathbb{E}^{Z_t^{(\mathfrak{D})}=x}\left[ g(Z_T^{(\mathfrak{D})}) - \int_t^T r(Z_s^{(\mathfrak{D})}, s)ds \right], \tag{25}$$

where $Z_t^{(\mathfrak{D})}, t \geq 0$, is the Itô process satisfying the SDE

---

[10] See Duffie (2001, Section 6L) and Shreve (2004, Section 5.5.1) for options on stocks with dividends in the Black-Scholes-Merton market.

$$dZ_t^{(\mathfrak{D})} = -\mathfrak{D}_t dt + v\big(Z_t^{(\mathfrak{D})}, t\big) d\tilde{B}(t), t \geq 0. \tag{26}$$

Next, we consider Bachelier's security $\mathcal{X}$ paying a cumulative dividend $\mathcal{D}_t, t \in [0, T]$, that is, the total amount of dividends paid until time $T$. The security price process is $X_t, t \in [0, T]$. The pair $(\mathcal{D}_t, X_t)_{t \in [0,T]}$ is defined on $(\Omega, \mathcal{F}, \mathbb{F}, \mathbb{P})$ and is called the **dividend-price pair**[11]. If $\mathcal{D}_t = \int_0^t \mathfrak{D}_s ds$, then $\mathfrak{D}_t, t \in [0, T]$, is called a dividend-rate process. The **gain process** is defined as $G_t = X_t + D_t$. Suppose that $\mathbb{Q}$ is the equivalent martingale measure $\mathbb{Q} \sim \mathbb{P}$ such that on $\mathbb{Q}$, $G_t - \beta_t = E_t^{(\mathbb{Q})}(G_T - \beta_T), t \in [0, T]$. This leads to Bachelier's security pricing formula[12]:

$$X_t = E_t^{(\mathbb{Q})}\left(X_T + \int_t^T dD_s - \int_t^T r_s ds\right), t \in [0, T]. \tag{27}$$

Consider a Bachelier's zero-coupon bond, that is, a security paying 1 at time $\tau \in [0, T]$. Thus, $D_t = 0$ for $t \in [0, \tau)$ and $D_t = 1$ for $t \in [\tau, T]$. According to (27), the value at $t \in [0, T]$ of the zero-coupon bond, denoted as $\mathcal{B}(t, \tau)$, is

$$\mathcal{B}(t, \tau) = E_t^{\mathbb{Q}}\big(1 - \int_t^\tau r_s ds\big) = 1 - E_t^{\mathbb{Q}}\big(\int_t^\tau r_s ds\big) \text{ for } t \in [0, \tau], \tag{28}$$

and $\mathcal{B}(t, \tau) = 0$ for $t \in (\tau, T]$. The collection $\{\mathcal{B}(t, \tau), 0 \leq t \leq \tau \leq T\}$ is called Bachelier's term structure of interest, and we will study it in more detail in the following section.

## 3. Bachelier's Term Structure of Interest Rates

### 3.1. Bachelier's Zero-coupon Bond Price

Consider BMM $(\mathcal{A}, \mathfrak{B}, \mathcal{C})$, where $\mathcal{C} = \mathcal{Z}_T$ is a zero-coupon bond[13] with a maturity $T > 0$, the asset price $A_t, t \geq 0$, is given by (1), and the simple interest account is given by (3). $\mathcal{Z}_T$ pays 1 at $T$ and nothing before that time; thus, **Bachelier's zero-coupon bond price $\mathcal{B}(t, T)$** of $\mathcal{Z}_T$ at time $t \in [0, T]$ is given by (18) or, equivalently, by (28):

$$\mathcal{B}(t, T) = E_t^{\mathbb{Q}}\left(1 - \int_t^T r_s ds\right) = 1 - E_t^{\mathbb{Q}}\left(\int_t^T r_s ds\right)[14]. \tag{29}$$

Let us determine a self-financing portfolio $P_t, t \in [0, T]$, of $\mathcal{A}$ and $\mathfrak{B}$ that can replicate $\mathcal{B}(t, T), t \in [0, T]$. According to (29), $\mathcal{B}(t, T) - \beta_t = E_t^{\mathbb{Q}}(1 - \beta_T)$ is a martingale on $\mathbb{Q}$ and thus,

$$\mathcal{B}(t, T) - \beta_t = E_0^{\mathbb{Q}}(1 - \beta_T) + \int_0^t \gamma_\mathcal{B}(s) dB_s^{(\mathbb{Q})}. \tag{30}$$

From (30), $d\mathcal{B}(t, T) = r_t dt + \gamma_\mathcal{B}(t) dB_t^{(\mathbb{Q})}$. For the self-financing portfolio $P_t = a_t A(t) + b_t \beta_t$, we have $P_t = (a_t + b_t) r_t dt + a_t v_t dB_t^{(\mathbb{Q})}$. As $dP_t = d\mathcal{B}(t, T)$, it follows that $a_t = \frac{\gamma_\mathcal{B}(t)}{v_t}$ and $b_t = 1 - \frac{\gamma_\mathcal{B}(t)}{v_t}$.

Given $\mathcal{T} > 0$, the set $\Lambda_\mathcal{T} = \{\mathcal{B}(t, T), 0 \leq t \leq T \leq \mathcal{T}\}$ will be called **Bachelier's term-structure of interest rates (BTSIR)** in $[0, \mathcal{T}]$.

### 3.2. Bachelier's Forward Contract

---

[11] We follow the exposition of arbitrage pricing with dividends given in Duffie (2001, Section 6L).

[12] The pricing formula (24) is the analogue of the security pricing formula in the BSMMM; see Duffie (2001, p. 125, formula (20)).

[13] See Duffie (2001, Section 6M) and Shreve (2004, Section 5.6.1). When we introduce and study Bachelier's term structure of interest rates (BTSIR), we follow the exposition of the TSIR given in Shreve (2004, Chapter 10) and Chalasani and Jha (1997).

[14] In the BSMMM, $\mathcal{B}(t, T) = E_t^{\mathbb{Q}}\left(exp\left(-\int_t^T r_s ds\right)\right)$.

**Bachelier's forward contract** is a slight modification [15] of the classical forward contract in the Black-Scholes-Merton market. Suppose that two market agents (ℶ and ℷ) enter Bachelier's forward contract; ℶ takes a long position in the contract and ℷ takes a short position. When the contract reaches maturity, ℶ's payoff is $V^{(l)}(T, T) = A_T - F(t, T)$ and ℷ's payoff is $V^{(s)}(T, T) = -V^{(l)}(T, T)$, where $F(t, T)$ is the **Bachelier's $T$-forward (delivery) price** of the asset $\mathcal{A}$. At time $t$, when the contract was initiated, ℶ and ℷ agreed on the value of Bachelier's forward contract, $V(t, T) = V_t \geq 0$, that is, the amount ℷ will receive from ℶ. To determine the value of Bachelier's $T$-forward price $F(t, T) = F(t, T; V_t)$, we apply the risk-neutral valuation formula (18). According to (18),

$$V_t = V(t, T) = E_t^{\mathbb{Q}}\left(V(T, T) - \int_t^T r_s ds\right) = E_t^{\mathbb{Q}}\left(A_T - F(t, T) - \int_t^T r_s ds\right)$$

$$= E_t^{\mathbb{Q}}(A_T - \beta_T) - F(t, T) + \int_0^t r_s ds = A_t - F(t, T).$$

Thus,

$$F(t, T) = A_t - V_t = A_t - V(t, T)[16].\tag{31}$$

Let us determine the hedging strategy that ℷ should follow. ℷ agrees to pay $F(t, T) = F(t, T; V(t))$ for the asset with value $A_T$ at $T$. At $t$, ℷ receives $V_t$. At $s \in [t, T]$, $V(s, T) = A_s - F(t, T)$. By (28), $1 = \mathcal{B}(s, T) + E_s^{\mathbb{Q}}\left(\int_s^T r_u du\right)$, and thus,

$$V(s, T) = A_s - F(t, T) = A_s - F(t, T)\left(\mathcal{B}(s, T) + E_s^{\mathbb{Q}}\left(\int_s^T r_u du\right)\right)$$

$$= A_s - E_s^{\mathbb{Q}}\left(\int_s^T r_u du\right) - F(t, T)\mathcal{B}(s, T).$$

$$\tag{32}$$

At time $s \in [t, T)$, ℷ can hedge her short position in Bachelier's forward contract by taking a short position of $F(t, T)$ zero-coupon bonds, each of which has a value of $\mathcal{B}(s, T)$. The value of this position is $F(t, T)\mathcal{B}(s, T) = A_t - V_t$. ℷ has received $V_t$ from her short position in Bachelier's forward contract. With the amount $A_t$, ℷ buys the asset. At the terminal time $T$, ℷ's investment position has a value of $A_T - F(t, T)\mathcal{B}(T, T) = A_T - F(t, T)$. ℷ delivers the asset and receives $F(t, T)$.

### 3.3. Bachelier's Future Price

We now pass to the definition of Bachelier's future price. Futures contracts in Bachelier's market can be viewed as "mark to market" [17] in a sequence of forward contracts. Let $t_{i,n}, i = 0, \ldots, n, 0 = t_{0,n} < t_{1,n} < \cdots < t_{n,n} = T, n \in \mathcal{N} = \{1,2,\ldots\}$, be a sequence of trading times (trading instances). To enter this contract, the trader ⊐ should pay $\varphi(t_{i,n})(t_{i+1,n} - t_{i,n}) \geq 0$. In the classical setting of futures contracts, $\varphi(t_{i,n}) = 0$, allowing ⊐ to take an unlimited number of futures contracts. In reality, ⊐ will pay margin and maintenance fees[18] to the exchange where the futures contract is traded. Taking unbounded numbers of futures contracts is impossible. In general, the assumption $\varphi(t_{i,n})(t_{i+1,n} - t_{i,n}) > 0$ can be treated as the transaction cost, and our analysis can follow the exposition in Janecek and Shreve (2010).

---

[15] In the traditional definition of a forward contract in the Black-Scholes-Merton market, it is assumed that at the time of contract initiation, the value of the forward contract is zero. Therefore, the determination of the forward (delivery) price is based on classical no-arbitrage assumptions; see Whaley (2012). Assuming that the value of the forward contract at its initiation is non-zero is not merely a simple transformation of the formula for the delivery price. If we assume that the forward contract is zero at its initiation, this means that traders entering the contract can multiply the value of the contract by any number, increasing the risk of their position without "paying for it." In other words, assuming that the value of the forward contract at its initiation is zero is unrealistic in business practice.

[16] According to (31), if Bachelier's forward contract value at time $t$, $V^{(0)}(t, T)$, is set to zero, that is, $V^{(0)}(t, T) = A_t - F^{(0)}(t, T) = 0$, then Bachelier's forward (delivery) price should be $F^{(0)}(t, T) = A_t$.

[17] See Tuovila (2022).

Following the definition of the future price in the Black-Scholes-Merton market[19], we call the $\mathbb{F}$-adapted process $\Phi_t, t \in [0,T]$, on $(\Omega, \mathcal{F}, \mathbb{F}, \mathbb{P})$ Bachelier's future price process, with an underlying asset $\mathcal{A}$, if (and only if)

$$\Phi_t = E_t^{\mathbb{Q}}(A_T), t = [0,T]. \tag{33}$$

According to (17), $\Phi_t = E_t^{\mathbb{Q}}(A_T) = A_t + \int_t^T r_u du$. Note that the discrete-time interpretation of the future in the BSMMM differs from a similar interpretation in BMM. Thus, here, we take (33) as the definition of Bachelier's future price process. In other words, Bachelier's future price $\Phi_t$ is the best forecast of the risk-neutral investor of the asset price $A_T$ at maturity $T$. To show the connection between Bachelier's forward contracts in discrete trading instances $t_{i,n}, i = 0, \ldots, n$, and Bachelier's future price, which is given in (32), we recall the formula for Bachelier's forward contract (31). Then,

$$V(t,T) = A_t - F(t,T). \tag{34}$$

For $t = t_{i,n}$ and $T = t_{i+1,n}$, (34) implies that $V(t_{i,n}, t_{i+1,n}) = A_{t_{i,n}} - F(t_{i,n}, t_{i+1,n})$. If we set $V(t_{i,n}, t_{i+1,n}) = \Phi_{t_{i,n}} = E_{t_{i,n}}^{\mathbb{Q}}(A_T)$, this will require that the delivery price $F(t_{i,n}, t_{i+1,n}) = F_\Phi(t_{i,n}, t_{i+1,n})$ is given by

$$F_\Phi(t_{i,n}, t_{i+1,n}) = \Phi_{t_{i,n}} - A_{t_{i,n}} = E_{t_{i,n}}^{\mathbb{Q}}(A_T) - A_{t_{i,n}} = E_{t_{i,n}}^{\mathbb{Q}}\left(A_T - A_{t_{i,n}}\right). \tag{35}$$

Thus, we can view Bachelier's future price $\Phi_{t_{i,n}}, i = 0, \ldots, n-1$, as Bachelier's forward contract $V(t_{i,n}, t_{i+1,n})$ with Bachelier's forward (delivery) price $F_\Phi(t_{i,n}, t_{i+1,n}) = E_{t_{i,n}}^{\mathbb{Q}}\left(A_T - A_{t_{i,n}}\right)$.

According to (34), the holder of a long Bachelier's forward contract initiated at $t = 0$ in which no payment is made at $t = 0$ will pay $F(0,T) + V(0,T)$ to receive the asset with price $A_T$. In comparison, the holder (⊐) of a long Bachelier's forward contract initiated at $t = 0$ who keeps the contract until maturity will pay $\Phi_T = A_T$ at maturity to receive the asset with price $A_T$. According to (35) and passing to the limit as $n \uparrow \infty$, ⊐ will collect a total payment of $\int_0^T d\Phi_t = \Phi_T - \Phi_0 = A_T - E^{\mathbb{Q}}(A_T)$. Thus, to receive the asset with value $A_T$, ⊐ has paid $\Phi_0 = E^{\mathbb{Q}}(A_T) = A_0 + \int_0^T r_u du$. Consider the spread $\Phi_0 - (F(0,T) + V(0,T))$ of the payments made by the future and forward holders to receive the same asset with value $A(T)$ at maturity $T$. According to (31) and (33),

$$\Phi_0 - \left(F(0,T) + V(0,T)\right) = A_0 + \int_0^T r_u du - A_0 = \int_0^T r_u du \in R.$$

If $\int_0^T r_u du > 0$, then ⊐ must make payments over all positive simple riskless rates, and for this reason, $\Phi_0 > \left(F(0,T) + V(0,T)\right)$. As $\Phi_0 = E^{\mathbb{Q}}(A_T) = A_0 + \int_0^T r_u du$ and $E^{\mathbb{P}}(A_T) =$ $A_0 + \int_0^T \rho_u du$, if $\Phi_0 < E^{\mathbb{P}}(A_T)$, then Bachelier's market is in normal backwardation[20]. If $\Phi_0 > E^{\mathbb{P}}(A_T)$, then Bachelier's market is in contango[21].

---

[18] See Scott (2021).

[19] See Shreve (2004, Definition 5.6.4, p. 244).

### 3.4. Bachelier's Forward Rates and Heath-Jarrow-Morton Model in Bachelier's Market Model

We start with the definition of Bachelier's forward agreement[22], which will allow us to determine an alternative representation of BTSIR:

$$\Lambda_{\mathcal{T}} = \Lambda_{\mathcal{T}}^{(r)} = \left\{ \mathcal{B}(t,T) = E_t^{\mathbb{Q}} \left( 1 - \int_t^T r_s ds \right), 0 \le t \le T \le \mathcal{T} \right\}. \tag{36}$$

At $t \ge 0$, consider borrowing 1 (\$1, for example) at time $T > t$, with a repayment of $\gamma(t,T,T+\varepsilon) = 1 + \varepsilon R(t,T,T+\varepsilon)$ at $T + \varepsilon$. This is equivalent to determining the number $\gamma(t,T,T+\varepsilon)$ of Bachelier's zero-coupon bonds $\mathcal{B}(t,T+\varepsilon)$, so that $\mathcal{B}(t,T) = \gamma(t,T,T+\varepsilon)\mathcal{B}(t,T+\varepsilon)$. Thus, $\gamma(t,T,T+\varepsilon) = 1 + \varepsilon R(t,T,T+\varepsilon) = \frac{\mathcal{B}(t,T)}{\mathcal{B}(t,T+\varepsilon)}$ and $\mathcal{R}(t,T,T+\varepsilon) = -\frac{\mathcal{B}(t,T+\varepsilon) - \mathcal{B}(t,T)}{\varepsilon \mathcal{B}(t,T+\varepsilon)}$. We call

$$f\!\!f(t,T) = \lim_{\varepsilon \downarrow 0} R(t,T,T+\varepsilon) = -\frac{1}{\mathcal{B}(t,T)} \frac{\partial \mathcal{B}(t,T)}{\partial T} = -\frac{\partial ln\mathcal{B}(t,T)}{\partial T} \tag{37}$$

**Bachelier's forward rate**, which represents the simple interest rate agreed upon at time $t \ge 0$ for borrowing and repaying the instantaneous loan at $T \ge t$. Thus, we obtain an alternative representation of BTSIR,

$$\Lambda_{\mathcal{T}}^{(f\!\!f)} = \left\{ \mathcal{B}(t,T) = \exp\left( -\int_t^T f\!\!f(t,u)du \right), 0 \le t \le T \le \mathcal{T} \right\}, \tag{38}$$

in terms of Bachelier's forward rates $f\!\!f(t,u), 0 \le t \le u \le \mathcal{T}$.

From (36), $\frac{\partial \mathcal{B}(t,T)}{\partial T} = -E_t^{\mathbb{Q}}(r_T)$ and $\frac{\partial \mathcal{B}(t,T)}{\partial T}\Big|_{T=t} = -r_t$. From (3.10), $\frac{\partial \mathcal{B}(t,T)}{\partial T} = -f\!\!f(t,T) \exp\left( -\int_t^T f\!\!f(t,u)du \right)$ and $\frac{\partial \mathcal{B}(t,T)}{\partial T}\Big|_{T=t} = -f\!\!f(t,t)$. Thus, $f\!\!f(t,t) = r_t$ for all $t \in [0,\mathcal{T}]$.

Next, we derive the dynamics of $\mathcal{B}(t,T), 0 \le t \le T \le \mathcal{T}$, in BTSIR by applying the HJM model for the TSIR[23]. We start with the stochastic dynamics of $f\!\!f(t,T), 0 \le t \le T \le \mathcal{T}$, on $(\Omega, \mathcal{F}, \mathbb{F}, \mathbb{P})$ given by the Itô process

$$df\!\!f(t,T) = \mu_{f\!\!f}(t,T)dt + \sigma_{f\!\!f}(t,T)dB_t, f\!\!f(0,T) \in R^{24}. \tag{39}$$

From (38) and since $f\!\!f(t,t) = r_t$,

$$d\mathcal{B}(t,T) = \left( r_t - \mu_{f\!\!f}^*(t,T) + \frac{1}{2}\sigma_{f\!\!f}^*(t,T)^2 \right) \mathcal{B}(t,T)dt - \sigma_{f\!\!f}^*(t,T)\mathcal{B}(t,T)dB_t, \tag{40}$$

where $\mu_{f\!\!f}^*(t,T) = \int_t^T \mu_{f\!\!f}(t,u)du$ and $\sigma_{f\!\!f}^*(t,T) = \int_t^T \sigma_{f\!\!f}(t,u)du$.

We search for the unique equivalent martingale measure $\mathbb{Q} \sim \mathbb{P}$ such that on $\mathbb{Q}$,

$$d\mathcal{B}(t,T) = r_t \mathcal{B}(t,T)dt - \sigma_{f\!\!f}^*(t,T)\mathcal{B}(t,T)B_t^{(\mathbb{Q})}, \tag{41}$$

---

[20] For normal backwardation in the BSMMM, see Hull (2012, p. 123 and p. 805) and Harper (2022).

[21] For contango in the BSMMM, see Hull (2012, p. 123 and p. 795).

[22] We follow the definitions of forward rates and the Heath-Jarrow-Morton (HJM) model for the TSIR in the BSMMM given in Shreve (2004, Section 10.3), adapting them for BTSIR.

[23] See Heath et al. (1992) and Jarrow (2002).

[24] $\mu_{f\!\!f}(t,T), \sigma_{f\!\!f}(t,T), t \ge 0$, are $\mathbb{F}$-adapted processes, and $\mu_{f\!\!f} : \mathbb{R} \times [0,T] \to \mathbb{R}$ and $\sigma_{f\!\!f} : \mathbb{R} \times [0,T] \to (0,\infty)$ satisfy the usual regularity conditions; see Duffie (2001, Appendix E).

where $B_t^{(\mathbb{Q})}$ is a standard Brownian motion on $\mathbb{Q}$. On $\mathbb{P}$, $B_t^{(\mathbb{Q})}$ is an arithmetic Brownian motion $dB_t^{(\mathbb{Q})} = dB_t + \theta_t dt$. Then, assuming that there exists $\theta_t, 0 \leq t \leq T$, independent of $T$ such that $\theta_t = \frac{\mu_f^*(t,T) - \frac{1}{2}\sigma_f^*(t,T)^2}{\sigma_f^*(t,T)}$ for all $0 \leq t \leq T$, then (41) holds. The dynamics of $f\!\!\!/(t,T), 0 \leq t \leq T \leq \mathcal{T}$, on $(\Omega, \mathcal{F}, \mathbb{F}, \mathbb{Q})$ are given by

$$d f\!\!\!/(t,T) = \sigma_{f\!\!\!/}(t,T)\sigma_f^*(t,T)dt + \sigma_{f\!\!\!/}(t,T)dB_t^{(\mathbb{Q})}. \tag{42}$$

The derivation of (41) and (42) is the same as in the classical HJM market model. This is disappointing, as $\mathcal{B}(t,T) = E_t^{\mathbb{Q}}\left(1 - \int_t^T r_s ds\right)$, while in the HJM model, the zero-coupon bond price is $B(t,T) = E_t^{\mathbb{Q}}\left(\exp\left(-\int_t^T \tilde{r}_s ds\right)\right)$, where $d\tilde{\beta}_t = \tilde{r}_t \tilde{\beta}_t dt, \tilde{\beta}_0 > 0$, represents the riskless bank account dynamics. Obviously, the stochastic dynamics of $\mathcal{B}(t,T)$ and $B(t,T)$, though they may look similar, are different. The reason that the derivation of (41) and (42) is the same as in the HJM model is the choice of the forward rates as the starting point in modeling BTSIR. From (37), it is clear that the definition of $f\!\!\!/(t,T)$ is not going to change if we replace $\mathcal{B}(t,T)$ with $\tilde{\mathcal{B}}(t,T) = E_t^{\mathbb{Q}}\left(\exp\left(-\int_t^T r_s ds\right)\right)$. While $\tilde{\mathcal{B}}(t,T)$ is close to $\mathcal{B}(t,T) = E_t^{\mathbb{Q}}\left(1 - \int_t^T r_s ds\right)$, they are different. With this consideration in mind, we introduce **no-interest loan rates** $\ell(t,T), 0 \leq t \leq T \leq \mathcal{T}$, as an alternative to Bachelier's forward rates $f\!\!\!/(t,T)$. Choosing the stochastic dynamics of $\ell(t,T), 0 \leq t \leq T \leq \mathcal{T}$, as a starting point in the analysis of BTSIR will lead to an alternative to the representation of BTSIR given by (36) and (38), namely,

$$\Lambda_{\mathcal{T}}^{(\ell)} = \left\{\mathcal{B}(t,T) = 1 - \int_t^T \ell(t,u)du, 0 \leq t \leq T \leq \mathcal{T}\right\}. \tag{43}$$

We call $\Lambda_{\mathcal{T}}^{(\ell)}$ the **Bachelier-Heath-Jarrow-Morton (BHJM) model** for the TSIR.

Let $T > t \geq 0$ and $t_k = t + k\Delta, k = 0, \ldots, n, n\Delta = T$. At time $t$, consider a **zero-simple-interest-loan** (shortly, no-interest loan) [25] $\mathcal{C}_k$ in which 1 ($\$1$, for example) is received at $t_k$ and 1 is paid back at $t_{k+1}$. At $t$, the cost $\ell(t, t_k; \Delta)$ for the provider of the loan $\mathcal{C}_k$ [26] is

$$\ell(t, t_k; \Delta)\Delta = \mathcal{B}(t, t_k) - \mathcal{B}(t, t_{k+1}) = E_t^{\mathbb{Q}}\left(\int_{t_k}^{t_{k+1}} r_s ds\right).$$

Thus, $\sum_{k=0}^{n-1} \ell(t, t_k; \Delta)\Delta = E_t^{\mathbb{Q}}\left(\int_0^T r_s ds\right) \sim \sum_{k=0}^{n-1} E_t^{\mathbb{Q}}(r_{t_k})\Delta$. By passing to the limit as $n \uparrow \infty, \Delta = \frac{T}{n} \downarrow 0$, we determine the no-interest loan rates:

---

[25] The no-interest loan is also known as a "non-interest-bearing loan" or a "zero-interest loan." In this type of agreement, the borrower receives a certain amount of money (in this case, $\$1$) at $t_k$ and is required to repay the same amount without any additional interest or charges at a specified future time (in this case, at time $t_{k+1}$). For some examples of zero-interest loans provided by government agencies to support certain types of purchases or investments, see Bonta (2023) and Service-Public.fr (2022). For no-interest loans offered to individuals, see Depersio (2023): *"Zero-interest loans, where only the principal balance must be repaid, often lure buyers into impulsively buying cars, appliances, and other luxury goods. These loans saddle borrowers with rigid monthly payment schedules and lock them into hard deadlines by which the entire balance must be repaid. Borrows who fail to honor the loan terms are subject to stiff penalties. These loans are typically only available to prospective buyers with FICO scores of 740 or higher."*

[26] This cost is typically paid by the receiver of the loan in the form of fees to the loan provider; the following is from Brozic (2022): *"No-interest loans can provide extra cash to pay a bill or cover an unexpected expense. But interest-free doesn't necessarily mean no cost. It's important to understand what fees — in addition to the principal — you may need to pay when getting a no-interest loan. We've rounded up our top picks with features such as low fees, access to money management tools, flexible repayment terms and the ability to build credit."*

$$\ell(t,T) = E_t^{\mathbb{Q}}(r_T), 0 \leq t \leq T \leq \mathcal{T}, \quad \int_t^T \ell(t,u)du = E_t^{\mathbb{Q}}\left(\int_t^T r_s ds\right), \ 0 \leq t \leq T \leq \mathcal{T}. \quad (44)$$

Thus, $\mathcal{B}(t,T) = E_t^{\mathbb{Q}}\left(1 - \int_t^T r_s ds\right) = 1 - \int_t^T \ell(t,u)du.$ We obtained an alternative representation of BTSIR, $\Lambda_{\mathcal{T}}^{(\ell)}$ (see (43)), in terms of the no-interest loan rates $\ell(t,T)$, $0 \leq t \leq T \leq \mathcal{T}$.

From (36), $\frac{\partial \mathcal{B}(t,T)}{\partial T} = -E_t^{\mathbb{Q}}(r_T)$ and $\frac{\partial \mathcal{B}(t,T)}{\partial T}\Big|_{T=t} = -r_t$. Similarly, from (43), $\frac{\partial \mathcal{B}(t,T)}{\partial T} = -\ell(t,T)\exp\left(-\int_t^T \ell(t,u)du\right)$ and $\frac{\partial \mathcal{B}(t,T)}{\partial T}\Big|_{T=t} = -l(t,t)$. Thus, $\ell(t,t) = r_t$ for all $t \in [0,\mathcal{T}]$.

Next, we derive the dynamics of $\mathcal{B}(t,T), 0 \leq t \leq T \leq \mathcal{T}$, in BTSIR (43) by applying risk-neutral valuation, like that in (38)–(42). We start with the stochastic dynamics for $\ell(t,T), 0 \leq t \leq T \leq \mathcal{T}$:

$$d\ell(t,T) = \mu_\ell(t,T)dt + \sigma_\ell(t,T)dB_t, \ell(0,T) \in R^{2}[27]. \quad (45)$$

As $\mathcal{B}(t,T) = 1 - \int_t^T \ell(t,u)du$, $d\mathcal{B}(t,T) = \left(r_t - \mu_\ell^{(*)}(t,T)\right)dt - \sigma_\ell^{(*)}(t,T)dB_t,$ where $\mu_\ell^{(*)}(t,T) = \int_t^T \mu_\ell(t,u)du$ and $\sigma_\ell^{(*)}(t,u) = \int_t^T \sigma_\ell(t,u)du.$ We search for the unique equivalent martingale measure $\mathbb{Q} \sim \mathbb{P}$ such that on $\mathbb{Q}$,

$$d\mathcal{B}(t,T) = r_t dt - \sigma_\ell^{(*)}(t,T)dB_t^{(\mathbb{Q})}, \quad (46)$$

where $B_t^{(\mathbb{Q})}$ is a standard Brownian motion on $\mathbb{Q}$. On $\mathbb{P}$, $B_t^{(\mathbb{Q})}$ is an arithmetic Brownian motion $dB_t^{(\mathbb{Q})} = dB_t + \theta_t^{(\ell)}dt$. Then,

$$d\mathcal{B}(t,T) = \left(r_t - \mu_\ell^{(*)}(t,T) + \sigma_\ell^{(*)}(t,T)\theta_t^{(\ell)}\right)dt - \sigma_\ell^{(*)}(t,T)dB_t^{(\mathbb{Q})}.$$

Next, if there exists $\theta_t^{(\ell)}, 0 \leq t \leq T$, independent of $T$ such that $\theta_t^{(\ell)} = \frac{\mu_\ell^{(*)}(t,T)}{\sigma_\ell^{(*)}(t,T)}$ for all $0 \leq t \leq T$, then (46) holds. Because $\mu_\ell^{(*)}(t,T) = \sigma_\ell^{(*)}(t,T)\theta_t^{(\ell)}, 0 \leq t \leq T$, is equivalent to $\mu_\ell(t,u) = \sigma_\ell(t,u)\theta_t^{(\ell)}, 0 \leq t \leq T$, we have that $\theta_t^{(\ell)} = \frac{\mu_\ell(t,u)}{\sigma_\ell(t,u)}$. Thus, the risk-neutral dynamics of $\ell(t,T)$ are given by

$$d\ell(t,T) = \sigma_\ell(t,T)dB_t^{(\mathbb{Q})}. \quad (47)$$

Comparing (47) with (42), we see that the "unfortunate" drift term $\sigma_\ell(t,T)\sigma_\ell^*(t,T)[28]$ in (42) that led to the introduction of the LIBOR market model[29] is not present in (47). This simplifies the implementation of the BHJM model $\Lambda_{\mathcal{T}}^{(\ell)}$. We begin the implementation of the BHJM model by estimating $\sigma_\ell(t,T)$, $0 \leq t \leq T \leq \mathcal{T}$[30]. For the historical estimation, we choose $\sigma_\ell(t,T) = \sigma_\ell\ell(t,T)$, where the constant $\sigma_\ell > 0$ is the volatility of the no-interest loan rates $\ell(t,T), 0 \leq t \leq T \leq \mathcal{T}$. We choose $\sigma_\ell$ to match historical data. Note that the diffusion coefficient $\sigma_\ell(t,T)$ is the same on $(\Omega, \mathcal{F}, \mathbb{F}, \mathbb{Q})$ and $(\Omega, \mathcal{F}, \mathbb{F}, \mathbb{P})$. By (47),

$$d\ell(t,T) = \sigma_\ell\ell(t,T)dB_t^{(\mathbb{Q})}, \quad (48)$$

---

[27] $\mu_\ell(t,T), \sigma_\ell(t,T), t \geq 0$, are $\mathbb{F}$-adapted processes, and $\mu_\ell: \mathbb{R} \times [0,T] \to \mathbb{R}$ and $\sigma_\ell: \mathbb{R} \times [0,T] \to (0,\infty)$ satisfy the usual regularity conditions; see Duffie (2001, Appendix E).

[28] See Shreve (2004, Section 10.4.1) and Chalasani and Jha (1997).

[29] The LIBOR market model is also known as the Brace-Gatarek-Musiela (BGM) model; see Brace et al. (1997) and Musiela and Rutkowski (2005, Section 12.4).

[30] We follow the exposition in Shreve (2004, Section 10.3.6) and Chalasani and Jha (1997).

where $\ell(0, T)$ is determined by the market data for non-interest-bearing loans. From (48), we determine $\ell(t, T)$ and $r_t = \ell(t, t), 0 \le t \le T \le \mathcal{T}$, on $(\Omega, \mathcal{F}, \mathbb{F}, \mathbb{Q})$. According to (46), we obtain the dynamics of $\mathcal{B}(t, T)$ on $(\Omega, \mathcal{F}, \mathbb{F}, \mathbb{Q})$:

$$d\mathcal{B}(t, T) = r_t dt - \sigma_\ell \left( \int_t^T \ell(t, u) \right) dB_t^{(\mathbb{Q})}, \qquad (49)$$

where $\mathcal{B}(0, T)$ is determined by market data. The dynamics of $\mathcal{B}(t, T)$ on $(\Omega, \mathcal{F}, \mathbb{F}, \mathbb{Q})$ are needed to value derivatives on the BTSIR $\Lambda_\mathcal{T}$. As $r_t = \ell(t, t), 0 \le t \le T \le \mathcal{T}$, on $(\Omega, \mathcal{F}, \mathbb{F}, \mathbb{Q})$ is also known, we can also determine the dynamics of the BTSIR $\Lambda_\mathcal{T} = \Lambda_\mathcal{T}^{(r)} = \left\{ \mathcal{B}(t, T) = E_t^\mathbb{Q} \left( 1 - \int_t^T r_s ds \right), 0 \le t \le T \le \mathcal{T} \right\}$ on the natural world $(\Omega, \mathcal{F}, \mathbb{F}, \mathbb{P})$.

Let $\tau = T - t \ge 0$ denote the time to maturity, $\mathcal{B}(t, T) = \mathcal{B}(t, t + \tau) = 1 - E_t^\mathbb{Q} \left( \int_t^{t+\tau} r_s ds \right) = \mathcal{D}(t, \tau)$, $r(t, \tau) = \ell(t, t + \tau)$, and $r(t, 0) = \ell(t, t) = r_t$. As $\mathcal{B}(t, T) = E_t^\mathbb{Q} \left( 1 - \int_t^T r_s ds \right) = 1 - \int_t^T \ell(t, u) du$ [31], we have an alternative representation for $\mathcal{D}(t, \tau)$:

$$\mathcal{D}(t, \tau) = \mathcal{B}(t, t + \tau) = E_t^\mathbb{Q} \left( 1 - \int_t^{t+\tau} r_s ds \right) = 1 - \int_t^{t+\tau} \ell(t, u) du. \qquad (50)$$

Thus, $\frac{\partial \mathcal{D}(t, \tau)}{\partial \tau} = \frac{\partial \mathcal{B}(t, t+\tau)}{\partial T} = -r(t, \tau)$. The notation $\sigma_\ell(t, T)$ can be replaced with $\sigma_\ell(t, \tau)$, and thus, (49) and (50) become

$$d\mathcal{B}(t, T) = r_t dt - \sigma_\ell^{(*)}(t, \tau) dB_t^{(\mathbb{Q})}, \ d\ell(t, T) = \sigma_\ell(t, \tau) dB_t^{(\mathbb{Q})}, \qquad (51)$$

where $\sigma_\ell^{(*)}(t, \tau) = \int_0^\tau \sigma_\ell(t, u) du$. Thus, by (51), $dr(t, \tau) = \frac{\partial r(t, \tau)}{\partial \tau} dt + \sigma_\ell(t, \tau) dB_t^{(\mathbb{Q})}$, and $d\mathcal{D}(t, \tau) = \left( r(t, 0) - r(t, \tau) \right) dt - \sigma_\ell^{(*)}(t, \tau) dB_t^{(\mathbb{Q})}$. As $\ell(t, t + \tau) = r(t, \tau) = \lim_{\delta \downarrow 0} \frac{1}{\delta} \int_\tau^{\tau+\delta} r(t, u) \, du$, we can write

$$\mathcal{L}(t, \tau; \delta) = \frac{1}{\delta} \int_\tau^{\tau+\delta} r(t, u) \, du. \qquad (52)$$

**Bachelier's forward LIBOR with tenor $\delta > 0$ [32].** The dynamics of $\mathcal{L}(t, \tau; \delta)$ are given by

$$d\mathcal{L}(t, \tau; \delta) = \frac{1}{\delta} \int_\tau^{\tau+\delta} dr(t, u) \, du$$

$$= \frac{1}{\delta} \left[ \left( r(\tau + \delta) - r(\tau) \right) dt + \left( \sigma_\ell^{(*)}(t, \tau + \delta) - \sigma_\ell^{(*)}(t, \tau) \right) dB_t^{(\mathbb{Q})} \right].$$

Furthermore, $\frac{\partial}{\partial \tau} \mathcal{L}(t, \tau; \delta) = \frac{1}{\delta} \frac{\partial}{\partial \tau} \int_\tau^{\tau+\delta} r(t, u) \, du = \frac{1}{\delta} \left( r(t, \tau + \sigma) - r(t, \tau) \right)$. Thus,

$$d\mathcal{L}(t, \tau; \delta) = \frac{\partial}{\partial \tau} \mathcal{L}(t, \tau; \delta) + \gamma(t, \tau; \delta) dB_t^{(\mathbb{Q})},$$

---

[31] We follow the exposition in Shreve (2004, Section 10.4), where the zero-coupon bond value is $B(t, T) = E_t^\mathbb{Q} \left( \exp \left( -\int_t^T \tilde{r}_s ds \right) \right)$, and $d\tilde{\beta}_t = \tilde{r}_t \tilde{\beta}_t dt, \tilde{\beta}_0 > 0$, represents the riskless bank account dynamics.

[32] In Shreve (2004, Chapter 10.4), the definition (52) is replaced by $L(t, \tau; \delta) = \frac{1}{\delta} \left[ \exp \left( \int_\tau^{\tau+\delta} r(t, u) \, du \right) - 1 \right]$, where $r(t, u) = f(t, t + u)$ and $f(t, t + u)$ is the forward rate. $L(t, \tau; \delta)$ is called the **forward LIBOR**, where $\delta > 0$ is the **tenor**; see Fernando (2021). The **LIBOR market model** is also known as the **Brace-Gatarek-Musiela (BGM) model**; see Brace et al. (1997) and Musiela and Rutkowski (2005, Section 12.4).

where $\gamma(t, \tau; \delta) = \frac{1}{\delta}\left(\sigma_\ell^{(*)}(t, \tau + \delta) - \sigma_\ell^{(*)}(t, \tau)\right)$ [33].

### 3.5. Hull and White Model in Bachelier's Market Model

The first definition of BTSIR (see (36)),

$$\Lambda_{\mathcal{T}} = \Lambda_{\mathcal{T}}^{(r)} = \left\{\mathcal{B}(t, T) = E_t^{\mathbb{Q}}\left(1 - \int_t^T r_s ds\right), 0 \le t \le T \le \mathcal{T}\right\}, \tag{53}$$

was in terms of the stochastic dynamics of the simple interest rate $r_t, t \in [0, \mathcal{T}]$, and this indicates that we should study BTSIR following the Hull and White (HW) model[34]. The HW model allows for negative interest rates, and this is important since negative interest rates are not exotic phenomena in the world's markets[35]. In 2023, Japan, Denmark, and Switzerland have negative interest rates (see Table 1).

**Table 1.** Countries with negative interest rates in 2023[36]

| Country | Interest Rate | 2023 Population |
|---|---|---|
| Japan | -0.01% | 123,294,513 |
| Denmark | -0.06% | 5,910,913 |
| Switzerland | -0.08% | 8,796,669 |

We start with the assumption that the simple interest rate $r_t, t \in [0, \mathcal{T}]$, in (53) has the following mean reverting dynamics on $(\Omega, \mathcal{F}, \mathbb{F}, \mathbb{Q})$:

$$dr_t = (a(t) - \mathcal{b}(t)r_t)dt + \mathcal{v}(t)dB_t^{(\mathbb{Q})}, t \in [0, \mathcal{T}], r_0 \in \mathbb{R}, \tag{54}$$

where $a(t), \mathcal{b}(t)$, and $\mathcal{v}(t), t \in [0, \mathcal{T}]$, are positive deterministic functions satisfying the standard regulatory conditions that guarantee the existence and strong uniqueness of the solution of (54). Then, setting $\mathcal{b}^{(*)}(t) = \int_0^t \mathcal{b}(u)du$, the solution of (54) is given by the Markov process

$$r_t = e^{-\mathcal{b}^{(*)}(t)}\left(r_0 + \int_0^t e^{\mathcal{b}^{(*)}(u)}a(u)du + \int_0^t e^{\mathcal{b}^{(*)}(u)}\mathcal{v}(u)dB_t^{(\mathbb{Q})}\right), t \in [0, \mathcal{T}]. \tag{55}$$

The dynamics of Bachelier's bond $\mathcal{B}(0, T) = E^{(\mathbb{Q})}\left(1 - \int_0^T r_t dt\right), 0 \le T \le \mathcal{T}$, are given by

$$\mathcal{B}(0, T) = 1 - E^{(\mathbb{Q})}\left(\int_0^T r_t dt\right) = 1 - \int_0^T e^{-\mathcal{b}^{(*)}(t)}\left(r_0 + \int_0^t e^{\mathcal{b}^{(*)}(u)}a(u)du\right)dt$$

$$= 1 - r_0 \mathbb{c}(0, T) - \mathbb{a}(0, T),$$

where $\mathbb{c}(0, T) = \int_0^T e^{-\mathcal{b}^{(*)}(v)}dv$ and $\mathbb{a}(0, T) = \int_0^T e^{-\mathcal{b}^{(*)}(t)}\left(\int_0^t e^{\mathcal{b}^{(*)}(u)}a(u)du\right)dt = \int_0^T e^{\mathcal{b}^{(*)}(v)}a(v)\left(\int_v^T e^{-\mathcal{b}^{(*)}(u)}du\right)dv$. Because $r_t, t \in [0, \mathcal{T}]$, is a Markov process,

---

[33] The implementation of $\mathcal{L}(t, \tau; \delta)$ in (51) is similar to that in Shreve (2004, p. 442), where $\gamma(t, \tau; \delta) = \frac{1 + \delta L(t, T; \sigma)}{\delta L(t, T; \delta)}\left(\sigma_\ell^{(*)}(t, \tau + \delta) - \sigma_\ell^{(*)}(t, \tau)\right)$.

[34] See Hull and White (1990, 1993). In our exposition of the Hull and White model in Bachelier's market model, we follow Shreve (2004, Section 6.5) and Chalasani and Jha (1997).

[35] See Haksar and Kopp (2020) and Foster (2020).

[36] See Neufeld (2022).

$$\mathcal{B}(t,T) = 1 - E^{(\mathbb{Q})}\left(\int_t^T r_u du\right) = 1 - r_t \mathbb{c}(t,T) - \mathbb{a}(t,T), \tag{56}$$

where

$$\mathbb{a}(t,T) = \int_t^T e^{\mathcal{b}(*)(v)} a(v)\left(\int_v^T e^{-\mathcal{b}(*)(u)} du\right) dv \text{ and } \mathbb{c}(t,T) = \int_t^T e^{-\mathcal{b}(*)(v)} dv. \tag{57}$$

We must determine the conditions on $a(t)$, $\mathcal{b}(t)$, and $v(t)$, $t \in [0,\mathcal{T}]$, that will ensure that BTSIR (53) is arbitrage-free. Consider the stochastic dynamics of $\mathcal{B}(t,T)$, $0 \le t \le T \le \mathcal{T}$:

$$d\mathcal{B}(t,T) = -dr_t \mathbb{c}(t,T) - d\mathbb{a}(t,T)$$

$$= -\left((a(t) - \mathcal{b}(t)r_t)dt + v(t)dB_t^{(\mathbb{Q})}\right)\mathbb{c}(t,T) - r_t \frac{\partial \mathbb{c}(t,T)}{\partial t}dt - \frac{\partial \mathbb{a}(t,T)}{\partial t}dt$$

$$= -\left[a(t) - \mathcal{b}(t)r_t + r_t\frac{\partial \mathbb{c}(t,T)}{\partial t} + \frac{\partial \mathbb{a}(t,T)}{\partial t}\right]dt - \mathbb{c}(t,T)v(t)dB_t^{(\mathbb{Q})}.$$

$$\tag{58}$$

According to (30), $d\mathcal{B}(t,T) = r_t dt + \gamma_{\mathcal{B}}(t)dB_t^{(\mathbb{Q})}$, and thus the HW model in Bachelier's market model is free of arbitrages if and only if

$$a(t) - \mathcal{b}(t)r_t + r_t\frac{\partial \mathbb{c}(t,T)}{\partial t} + \frac{\partial \mathbb{a}(t,T)}{\partial t} + r_t = 0,$$

for all $0 \le t \le T \le \mathcal{T}$.

Consider a European call option on Bachelier's zero-coupon price $\mathcal{B}(t,T)$, $t \in [0,T]$, with a maturity $T$. The maturity of the call option is $\tau \in (0,T)$, with a terminal value $(\mathcal{B}(\tau,T) - K)^+ = \max(\mathcal{B}(T_1,T) - K, 0)$, where $K \in \mathbb{R}$ is the strike price. According to (18), the value of the call option at $t = 0$ is

$$C_0 = E^{(\mathbb{Q})}((\mathcal{B}(\tau,T) - K)^+ - R_\tau)$$
$$= E^{(\mathbb{Q})}((1 - r_\tau\mathbb{c}(\tau,T) - \mathbb{a}(\tau,T), -K)^+ - R_\tau)$$
$$= \int_{-\infty}^{\infty}\int_{-\infty}^{\infty}[(1 - y\mathbb{c}(\tau,T) - \mathbb{a}(\tau,T), -K)^+ - x] f_{R_\tau,r_\tau}(x,y)dxdy,$$

where $f_{R_\tau,r_\tau}(x,y), x \in \mathbb{R}, y \in \mathbb{R}$, is the density of $R_\tau = \int_0^\tau r_s ds$ and $r_\tau$. The density $f_{R_\tau,r_\tau}$ is bivariate normal, with[37]

$$\mu_{R_\tau} = E^{(\mathbb{Q})}(R_\tau) = \int_0^\tau\left[e^{-\mathcal{b}(*)(v)}\left(r_0 + \int_0^v e^{\mathcal{b}(*)(u)} a(u)du\right)\right]dv,$$

$$\sigma_{R_\tau}^2 = E^{(\mathbb{Q})}(R_\tau - \mu_{R_\tau})^2 = \int_0^\tau\left[e^{2\mathcal{b}(*)(v)}v(v)^2\left(\int_v^\tau e^{-\mathcal{b}(*)(u)} du\right)^2\right]dv,$$

$$\mu_{r_\tau} = E^{(\mathbb{Q})}(r_\tau) = e^{-\mathcal{b}(*)(v)}\left(r_0 + \int_0^v e^{\mathcal{b}(*)(u)} a(u)du\right),$$

$$\sigma_{r_\tau}^2 = E^{(\mathbb{Q})}(r_\tau - \mu_{r_\tau})^2 = e^{-2\mathcal{b}(*)(v)}\int_0^\tau e^{2\mathcal{b}(*)(v)}v(v)^2 dv,$$

and a covariance

---

[37] See Chalasani and Jha (1997).

$$cov(R_\tau, r_\tau) = \int_0^\tau e^{-\mathcal{E}^{(*)}(u) - \mathcal{E}^{(*)}(\tau)} \left( \int_0^u e^{2\mathcal{E}^{(*)}(v)} \mathit{v}(v)^2 dv \right) du.$$

## 4. Conclusions

In this paper, we explore the fundamentals of dynamic asset pricing within Bachelier's market model (BMM), an extension of Louis Bachelier's groundbreaking option pricing model. We leverage the arithmetic Brownian motion within BMM to represent risky asset price dynamics, coupled with a simple interest bank account that represents the riskless asset. Our investigation delves into Bachelier's partial differential equation (PDE), the solution of which provides a formula for European contingent claims (ECCs). This approach highlights the fundamental differences between BMM and the Black-Scholes-Merton market model (BSMMM), and it allows the expansion of BMM to include assets yielding a simple dividend. We also probe the no-arbitrage price of zero-coupon bonds within BMM, which leads us to explore Bachelier's term structure of interest rates (BTSIR). We present a novel version of Bachelier's Heath-Jarrow-Morton (HJM) model, replacing forward rates with zero-simple-interest-loans, which introduces a new form of BTSIR. Lastly, the Hull-White (HW) interest rate model is adapted so that it aligns with BMM. The paper concludes with the investigation of BMM's applicability in cases such as those involving ESG-adjusted stock prices and commodity spreads, introducing new dimensions into traditional investment frameworks.

In future work, we plan to explore the valuation of ESG-adjusted prices for US equity on both the spot and derivative markets. Additionally, we intend to extend BMM to account for asset prices that follow Lévy processes, drawing upon the exposition in Schoutens (2003) and Shirvani et al. (2021).

**Author Contributions:** Conceptualization, S.T.R.; methodology, S.T.R.; formal analysis, S.T.R., N.A.N., B.D., G.J., B.O., P.Y.; investigation, S.T.R., N.A.N., B.D., G.J., B.O., P.Y.; writing—original draft preparation, S.T.R.; writing—review and editing, S.T.R., N.A.N., B.D., G.J., B.O., P.Y.; supervision, S.T.R.; project administration, S.T.R. All authors have read and agreed to the published version of the manuscript.

**Funding:** This research received no external funding.

**Conflicts of Interest:** The authors declare no conflict of interest.